\documentclass[aps,showpacs,floatfix,twocolumn,amsmath,amssymb,prd]{revtex4}
\usepackage[citecolor=blue,colorlinks=true,urlcolor=blue,linkcolor=blue]{hyperref}
\usepackage{tikz,graphicx}
\usepackage{slashed,dsfont,bm}
\usepackage{subfigure}
\usepackage{stmaryrd}
\usepackage{mathbbol}
\usepackage{bbold}
\usepackage{eepic}
\usepackage{pst-all}
\usepackage{pstricks-add}
\usepackage{lscape}
\usepackage{rotating}
\usepackage{multirow}
\usepackage{hhline}
\usepackage{soul}
\newcommand{\Ef}{\mathcal{E}}

\newcommand{\Ea}{\text{EN1}}

\newcommand{\Ec}{\text{EN2}}

\begin{document}

\def\fm {\,{\rm fm}}
\def\MeV {\,{\rm MeV}}
\def\GeV {\,{\rm GeV}}

\title{Electric polarizability of neutral hadrons from dynamical lattice QCD ensembles}
\author{M. Lujan, A. Alexandru, W. Freeman, and F.X. Lee \\
\vspace*{0.5cm} }
\affiliation{ Physics Department, The George Washington University, Washington, DC 20052, USA \hfill\\}

\begin{abstract}
\begin{center} {\bf{Abstract}} \end{center}
We present a valence calculation of the electric polarizability of the neutron, neutral pion,
and neutral kaon on two dynamically generated nHYP-clover ensembles. The pion masses for these ensembles are
227(2) MeV and 306(1) MeV, which are the lowest ones used in polarizability studies. This is part of a program geared 
towards determining these parameters at the physical point. We carry out a high statistics calculation that allows us to: 
(1)~perform an extrapolation of the kaon polarizability to the physical point; we find $\alpha_K=0.269(43)\times10^{-4}$fm$^{3}$,
(2)~quantitatively compare our neutron polarizability results with predictions from $\chi$PT, and
(3)~analyze the dependence on both the valence and sea quark masses. The kaon polarizability varies slowly with the
light quark mass and the extrapolation can be done with high confidence. 
\end{abstract} 
\pacs{12.38.Gc}

\maketitle

\section{Introduction}
\label{secintroduction}
Determining the polarizability of hadrons has been a challenge both theoretically and experimentally for several decades. 
To calculate them from first principles, one needs a nonperturbative approach to QCD. In this work we use lattice QCD.  The lattice calculation needs to overcome many challenges in order to get to experimentally relevant results, such as the use of smaller quark masses, determining an appropriate field strength for the electric field, volume effects, sea-quark charging effects, {\it{etc}}. In this paper, we will study several of these issues for neutral hadrons (neutron, neutral pion and kaon). 
Charged hadrons involve additional complications which we defer to future studies.

The neutron polarizability is known experimentally, $\alpha_n = 11.6(1.5)\times 10^{-4}\fm^3$ \cite{Beringer:1900zz}, and therefore a theoretical calculation from first principles provides a good test of QCD. The first experimental determination of the kaon polarizability will be performed as part of the COMPASS \cite{Abbon:2007pq,Nagel:1484476} experiment along with more precise determination of the pion polarizability. 

At the lowest order the effects of an electromagnetic field on hadrons can be parameterized by the effective Hamiltonian:
\begin{equation}
\mathcal{H}_{em} = -\vec{p}\cdot\vec{\Ef} -\vec{\mu}\cdot\vec{B} -\frac{1}{2}\left(\alpha \Ef^2 + \beta B^2\right)+...,
\end{equation}
where $p$ and $\mu$ are the static electric and magnetic dipole moments, respectively, and $\alpha$ and $\beta$ are the static electric and magnetic polarizabilities. Due to time reversal symmetry of the strong interaction, the static dipole moment, $\vec{p}$, vanishes.  Furthermore, by restricting ourselves to the case of a constant electric field, the leading contribution to the electromagnetic interaction comes from the electric polarizability term at $\mathcal{O}(\Ef^2)$.

There have been several studies on computing the electric polarizabilities in lattice QCD \cite{Engelhardt:2007ub,Detmold:2010ts,Detmold:2009dx,Fiebig:1988en,Alexandru:2008sj}.
These calculations, however, were done at  relatively large pion masses leaving the chiral region largely unexplored.
Here we present  a study using two flavors of dynamical nHYP-clover fermions with two different dynamical pion masses (227 and 306 MeV) and several partially quenched valence masses. These pion masses are the lowest pion masses to date for polarizability studies.  

Our calculation employs the background field method and uses Dirichlet boundary conditions (DBC) for the valence quarks. This choice of boundary condition has the benefit of allowing us to freely chose an arbitrarily small value for the electric field which is needed in order to extract the polarizability.  We note that this work, though done on dynamical configurations, uses electrically neutral sea quarks throughout. Methods for introducing the effect of the electric field on the sea quarks are under investigation~\cite{Engelhardt:2007ub, Freeman:2012cy}.

The paper is organized as follows. In Section~\ref{secmethodology} we describe our methods used in extracting the
polarizability. This includes a discussion of the background field method,  how the value of the electric field is chosen, and our fitting procedure. In Section~\ref{secresults} we present our results for the neutron, pion, and kaon. Section~\ref{sec:discussion} is a discussion on the chiral behavior of our results. Lastly we will conclude and outline some future studies in Section~\ref{sec:conclusion}. 

\bigskip

\section{ Methodology}
\label{secmethodology}

\subsection{ Background Field Method}

Electromagnetic properties of hadrons can be determined with the  background field method \cite{Martinelli:1982cb}.  This procedure introduces the electromagnetic vector potential, $A_{\mu}$, to the Euclidean QCD Lagrangian via minimal coupling; the covariant derivative becomes
\begin{equation}
D_{\mu} = \partial_{\mu} -igG_{\mu} -iqA_{\mu},
\end{equation}
where $G_{\mu}$ are the gluon field degrees of freedom. The lattice implementation  is achieved by a multiplicative U(1) phase factor to the gauge links {\it{i.e.}},
\begin{equation}
U_{\mu} \rightarrow e^{-iqaA_{\mu}} U_{\mu}.
\end{equation}

To achieve a constant electric field, say in the $x$-direction, we may choose $A_{x} = -i\Ef t$ which provides us with a real multiplicative factor.  The extra factor, $i$, comes from the fact that we are using a Euclidean metric on the lattice.  
A more convenient choice, and the one implemented in this study,  is the use of an imaginary value for the electric field which leads to a U(1) multiplicative factor that keeps the links unitary. When using an imaginary value of the field, the energy shift due to the polarizability acquires an additional negative sign leading to a positive energy shift for a positive value of the polarizability \cite{Alexandru:2008sj}:
\begin{equation}
\delta E = -\frac{1}{2} \alpha \Ef^2~ \rightarrow ~ \delta E = +\frac{1}{2} \alpha \Ef^2
\end{equation}

To compute this energy shift (and hence the polarizability)  we calculate the zero-field ($G_0$), plus-field ($G_{+\Ef}$),  and minus-field ($G_{-\Ef}$) two-point correlation functions for the interpolating operators of interest. The combination of the plus and minus field correlators allows us to remove any $\mathcal{O}(\Ef)$ effects, which are statistical artifacts, when the sea quarks are neutral. For neutral particles in a constant electric field the correlation functions still retain their single exponential decay in the limit $t \rightarrow \infty$. In particular we have
\begin{equation}
\lim_{t \to \infty} \langle G_{\Ef} \rangle = A(\Ef)\exp[-E(\Ef)~t], 
\label{eqn::corr}
\end{equation}
where $E(\Ef)$ has the perturbative expansion in the electric field given by
\begin{equation}
E(\Ef) = m + \frac{1}{2}\alpha \Ef^2 + ... ~.
\end{equation}
By studying the variations of the correlation functions with and without an electric field one can isolate the energy shift to obtain $\alpha$.

For spin-1/2 particles the energy $E(\Ef)$ has an additional contribution to its energy shift if the hadron has a magnetic moment, as with the neutron. The interaction of the magnetic moment with the external field contributes to the energy shift at the same order, $\mathcal{O}(\Ef^2)$, as the Compton polarizability \cite{Lvov:1993fp}. To see this consider a  point-like neutral spin-1/2 particle with a non-zero magnetic moment. This system satisfies the Pauli-Dirac equation (in Minkowski space) when subject to an electromagnetic field \cite{BurlingClaridge:1989zr}, in particular
\begin{equation}
 \left( i\gamma_{\mu} \partial^{\mu} - m - \frac{\mu}{2m}F_{\mu\nu}\sigma^{\mu\nu}\right)\psi =0,
\label{mueqn}
\end{equation}
where $F_{\mu \nu}$ is the electromagnetic field strength tensor. In the non-relativistic limit one can approximate the solution as 
\begin{equation}
\psi \approx e^{-i mt}\left( \begin{array}{c}
 \Phi \\
\chi \\
\end{array} \right).
\end{equation}
In the case of a constant electric field the equation of motion for the upper component satisfies the differential equation
\begin{equation}
i\frac{\partial\Phi}{\partial t} = \left[ \frac{p^2}{2m} - \vec{\mu} \cdot\left(\vec{\Ef} \times \frac{\vec{p}}{m} \right) + \frac{\mu^2}{2m} \Ef^2 \right] \Phi.
\label{mueqn2}
\end{equation}
The first term is the usual kinetic term associated with a particle of momentum $p$. The second term is very interesting; in general there is an $\mathcal{O}(\Ef)$ contribution to the energy shift due to the magnetic moment and momentum of the hadron. It would appear that this term would affect our energy shift since our system has a non-zero momentum due to the DBC (see discussion below).  However, this term is zero for our calculations since the direction of the electric field is in the same direction as its induced momentum. The last term is the non-zero contribution due to the magnetic moment of the particle. We see that not only is there a contribution to the energy shift from the Compton polarizability, but also a contribution from the neutron's magnetic moment. Note that the Compton polarizability acts in the opposite direction of the effects due to the magnetic moment. In the case of a real electric field it lowers the energy  whereas the magnetic moment effect increases it. For imaginary fields, as we use in this study, the effect is opposite. The energy expansion is then augmented as
\begin{equation}
E(\Ef) = m + \frac{1}{2}\Ef^2\left(\alpha_c- \frac{\mu^2}{m}\right) + ...~.
\label{eqn::Eneutron}
\end{equation}
This combination of the Compton polarizability term and the magnetic moment term is what is called the static polarizability,
\begin{equation}
\alpha = \alpha_c- \frac{\mu^2}{m},
\end{equation}
which is what we measure from the energy shift of the hadron. To obtain the Compton polarizability ($\alpha_c$) we need to account for the magnetic moment. We will address how this is done for the nucleon when presenting our results.

 \begin{figure}[t]
\includegraphics[width= 3.65 in]{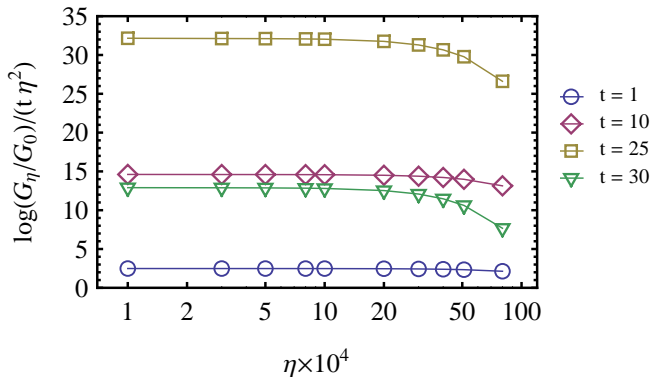}
\caption{Scaling check for the correlation function of  $\bar{d} \gamma_5 d$ on a single $\Ea$ gauge configuration. The constant behavior indicates the values of $\eta$  for which quadratic scaling holds. Deviations from this behavior are apparent for the larger values of $\eta$.}
\label{etascaling}
\end{figure}

To extract the energy shift we need to choose an appropriate value of $\Ef$. If the field is too large then higher order effects become non-negligble, and if the field is too small we will encounter numerical instabilities in the fitting procedure since the shift to be extracted becomes comparable to the numerical errors introduced by roundoff effects. To ascertain an appropriate field strength we take a single gauge configuration and compute 
\begin{equation}
\log\frac{G_{\eta}(t)}{G_0(t)}/(t\eta^2)
\end{equation} 
for the interpolating operator, $\bar{d} \gamma_5 d$, to calibrate the field as a function of $\eta \equiv a^2 q_d\Ef$ where $a$ is the lattice spacing and $q_d$ is the magnitude of the electric charge for the down quark. The correlator, $G_{\eta}(t)$, is symmetrized with respect to $\eta$ to ensure that there are no linear effects. This symmetrization is achieved by algebraically averaging the plus and minus field correlators. This function is proportional to the effective energy shift and should have a flat behavior in the region where quadratic scaling dominates. Deviations from a constant
behavior indicates effects coming from higher order terms in $\Ef$. Fig.~\ref{etascaling} shows our findings. We see the effects beyond $\mathcal{O}(\Ef^2)$ more dominant for larger times and larger fields.  The value $\eta=10^{-4}$ is in a well-behaved scaling region for the time slices that we use in our fits and therefore we use this field strength for all our calculations. We checked the $\mathcal{O}(\Ef^2)$ behavior for several different configurations on all ensembles used in this work and found similar trends as in Fig.~\ref{etascaling}.

Several studies used periodic boundary conditions (PBC) in their calculations \cite{Detmold:2010ts,Detmold:2009dx}. In this case, in order to produce a constant electric field, the value of the electric field must be quantized in units of $\Ef_0 = 2 \pi/(q T L)$ where $T$ and $L$ are the physical extent of the lattice in the temporal and spatial directions respectively. For the lattice sizes used in this study, the smallest values of $\eta$ which satisfy this quantization condition are $\eta = 0.005$ and $\eta = 0.004$ for the $24^3 \times 48$ and $24^3 \times 64$ ensembles respectively. These values of $\eta$ are on the periphery where quadratic scaling begins to break down.

\begin{figure}[b]
\centering
\includegraphics[width= 3.3in]{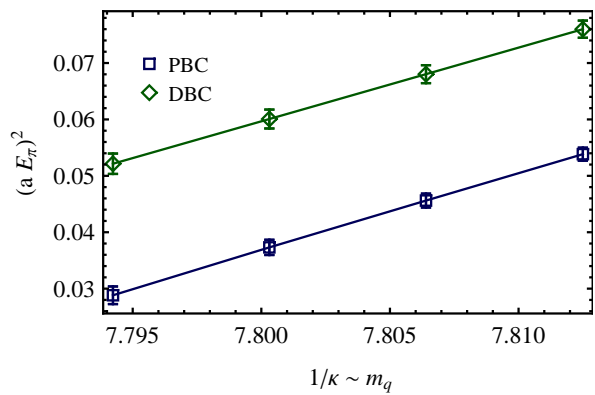}
\caption{Plot of $(a E_{\pi})^2$ as a function of the bare quark mass ($\sim 1/\kappa$) for DBC and PBC for the 306 MeV ensemble.}
\label{plot:pionbc}
\end{figure}

To free ourselves of this constraint, we use Dirichlet boundary conditions (DBC) in the $x$ direction. This allows us to deploy arbitrarily small values of the field. The use of DBC in the direction of the field introduces boundary effects, one of which is an induced momentum that vanishes in the limit $L\rightarrow \infty$. For an elementary particle in a box of length $L$ the induced momentum is $p\approx \pi/L$ and the lowest energy state is $ E = \sqrt{m^2 + p^2}$.  One can observe this effect by looking, for example, at the lowest energy in the pion channel with DBC and PBC.  This is shown in Fig.~\ref{plot:pionbc}.  Note the constant shift from the DBC to PBC that corresponds to $p^2$. Throughout this work we plot the results as a function of the pion mass, as a proxy for the quark mass. 
This is the pion mass measured using PBC.
 
When the hadron is moving, the energy shift ($\delta E$) induced by the electric field is not equal to the change in the hadron mass since $E = \sqrt{m^2 + p^2}$. We compute the mass shift, $\delta m$, as
\begin{equation}
\delta m = \delta E \frac{E}{m},
\end{equation}
where $m$ is the zero-momentum mass of the particle which we calculate using PBC.

Using the methods just described we aim to calculate the polarizability of the pion, kaon and neutron. Before proceeding, we should mention that we are only computing the connected contribution to the pion correlation function. The standard interpolating field for the neutral pion is $\pi^0 = \left (\bar{u}\gamma_5u - \bar{d}\gamma_5 d\right)/\sqrt{2}$.  When the theory is isospin symmetric the disconnected contributions to the two-point correlation function cancel. In the case of an applied external electric field we no longer have isospin symmetry and we should include the disconnected contributions to the correlator. This requires significantly more computational efforts and is thus neglected in this study. We will comment more on this issue in the discussion section.

\subsection{ Fitting Method}
\label{fitmethod}

The zero, plus, and minus-field correlation functions are highly correlated because they are constructed from the same gauge configurations. Extracting the energy shift, $\delta E$, then requires a simultaneous-correlated fit among the three correlators. We emphasize that this desired energy shift is very small, several orders of magnitude smaller than the statistical uncertainties of the energy itself. It is because of the correlation that we can extract this tiny shift. Thus, our only hope of extracting such a tiny shift lies in the ability to properly account for the correlations among the three values of the electric field. To do this we construct the following difference vector as
\begin{eqnarray}
\mathbf{v}_{i} &\equiv& f(t_i) - \langle G_{0}(t_i)\rangle,\\
\mathbf{v}_{N+i} &\equiv&  \bar{f}(t_{i}) - \langle G_{+\Ef}(t_{i})\rangle, \nonumber\\
\mathbf{v}_{2N+i} &\equiv&  \bar{f}(t_{i}) - \langle G_{-\Ef}(t_{i})\rangle~~~\mbox{for}~i=1,...,N \nonumber
\end{eqnarray}
where  $t_1...t_N$ is the fit window, $f(t) = A~e^{-E t}$ and $\bar{f}(t) =(A+\delta A)~e^{-(E+\delta E)t}$. We minimize the $\chi^2$ function, 
\begin{equation}
\chi^2= \frac{1}{2} \mathbf{v}^{\mathbf{\dagger}}~ \mathbf{C}^{-1}~ \mathbf{v}, \nonumber
\end{equation}
in the usual fashion, where $\mathbf{C}$ is the $3N~\times~3N$ correlation matrix which has the block structure
\[ \mathbf{C} = \left( \begin{array}{ccc}
C_{0 0} ~& C_{0 +}~ & C_{0 -} \\
C_{+ 0} ~& C_{+ +}~ & C_{+ -} \\
C_{- 0} ~& C_{- +} ~& C_{- -} \end{array} \right),\] 
where $0,+,-$ represent  $G_0, G_{+\Ef}$, and $G_{-\Ef}$ respectively. Note that the symmetrization is done implicitly in this procedure, since $\bar{f}$ is the same for  $G_{+\Ef}$, and $G_{-\Ef}$. This method is used to extract all parameters presented in this work.

\section{Results}
\label{secresults}

\subsection{ Ensemble Details}
\label{sec:ensemble}
The electric polarizabilities of the neutron, pion, and kaon are calculated on two dynamically generated 2-flavor nHYP-clover ensembles \cite{Hasenfratz:2007rf} with two different sea quark masses to study the chiral behavior. Details of the ensembles are listed in Table~\ref{tab:ensembles}. The lattice spacing for both ensembles was computed using the Sommer scale~\cite{Sommer:1993ce}. Our calculation of the Sommer parameter, $r_0/a$ follows the methodology described in \cite{Pelissier:2012pi}. We find $r_0/a$ to be 4.017(50) and 4.114(39) for ensembles $\Ea$ and $\Ec$, respectively. Using the value of $r_0 =0. 5$~fm yields the lattice spacings listed in Table~\ref{tab:ensembles}.
\begin{table}[t]
\begin{center}
\begin{tabular}{|c|c|c|c|c|c|c|}
\hline
Ensemble &Lattice&  $a$ (fm) &  $\kappa$& $m_{\pi}$ (MeV)& $N_{\text{c}}$& $N_{s}$\\ 
\hline
\hline
$\Ea$&$24^3 \times 48$& 0.1245(16)& 0.12820   & $ 306(1)$& 300&25\\
$\Ec$&$24^3 \times 64$& 0.1215(11)& 0.12838 &  $227(2)$& 450&18\\
\hline
\end{tabular}
\end{center}
\caption{Details of the lattice ensembles used in this work. $N_c$ and $N_s$ label the number of configurations and number of sources on each configuration, respectively.}
\label{tab:ensembles}
\end{table}

\begin{figure}[b]
\includegraphics[width= 3.3in,height=2.1in]{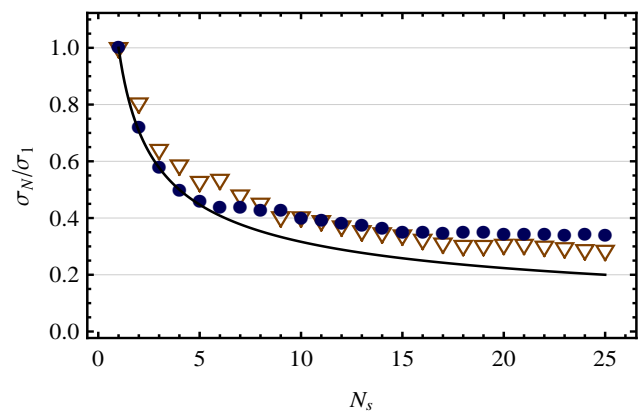}
\caption{Error scaling for the pion (blue/circle) and neutron (orange/triangle) for the $\Ea$ ensemble. The points correspond to the measured values of the uncertainty in the energy shift, $\delta E$. The solid black line corresponds to the expected ideal situation if the data points were completely uncorrelated. The curves were generated by taking the value for $N_s=1$ and scaling it by $1/\sqrt{N_s}$.}
\label{plot:sources}
\end{figure}

In addition to computing propagators for the sea quark mass we calculated propagators for a string of partially quenched quark masses to gauge the influence of the sea quark mass on the polarizability.  The partially quenched values were estimated by
looking at the dependence of $m_{\pi}$ on $\kappa$ for each ensemble. We then performed interpolation in 
the region where $m_{\pi}^2 \propto 1/\kappa$ to obtain values of $\kappa$ which gave reasonable values of $m_{\pi}$ around
the dynamical pion masses. The partially quenched $m_{\pi}$ and $\kappa$ values for each ensemble are tabulated in Table~\ref{tab:partialquench}. 
\begin{figure*}[t]
\includegraphics[width= 3.1in]{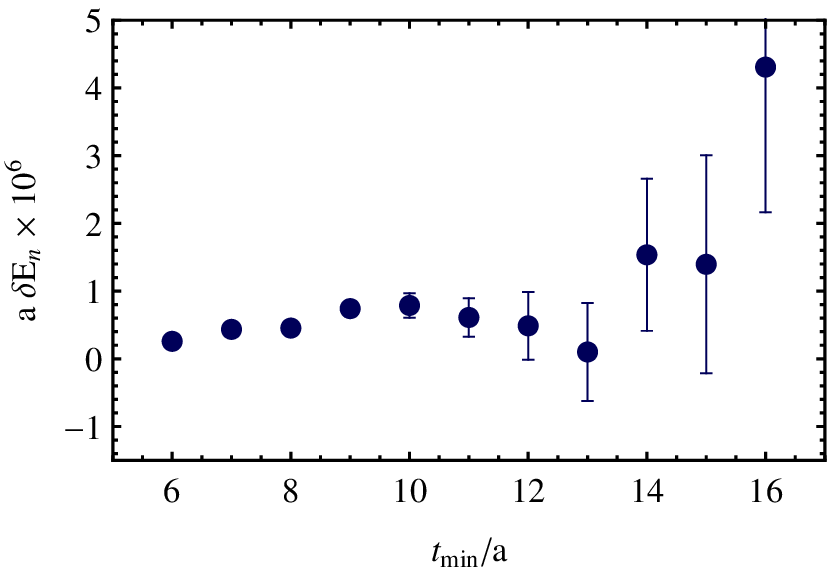}
\includegraphics[width= 3.1in]{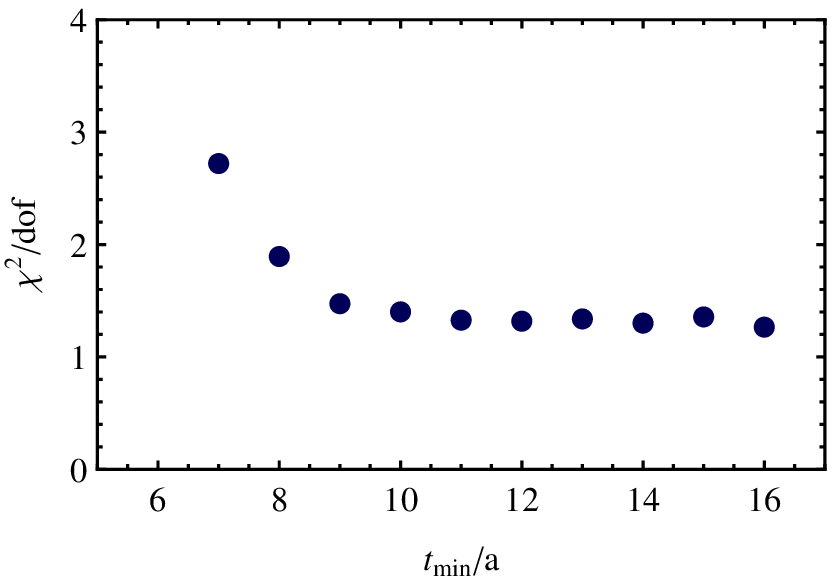}
\caption{Plots of the extracted values of $\delta E$ and $\chi^2/\text{dof}$ for the neutron on the $\Ec$ lattice as a function of minimum time window used in the fits.}
\label{plot:fitwindow}
\end{figure*}

\begin{table}[h]
\begin{center}
\begin{tabular}{|c|c|c|c|c|c|c|c|c|c|c|c|c|c|}
\hline
Ensemble& Quantity& \multicolumn{1}{|c|}{$m_1$} & \multicolumn{1}{|c|}{$m_2$} &\multicolumn{1}{|c|}{$m_3$}  &\multicolumn{1}{|c|}{$m_4$} &  \multicolumn{1}{|c|}{$m_5$} \\
 \hline
 \hline
\multirow{2}{*}{$\Ea$}&$m_{\pi}$ [MeV] &269(1)&\textcolor{black}{\bf{306(1)}}&339(1)&368(1)&---\\ 
                                       &$\kappa$ &0.1283 & {\bf{0.1282}} & 0.1281 & 0.1280 & --- \\ 
                                   \hline
\multirow{2}{*}{$\Ec$}&$m_{\pi}$ [MeV] &\textcolor{black}{\bf{227(2)}}&283(1)&322(1)&353(1)&382(1) \\
                                       &$\kappa$ &{\bf{0.12838}}& 0.12825& 0.12814& 0.12804& 0.12794\\ 
\hline
\end{tabular}
\end{center}
\caption{List of pion masses and kappa values used for each ensemble. The boldface values correspond to the fully dynamical (or unitary) points.}
\label{tab:partialquench}
\end{table}
An optimally implemented multi-GPU Dslash operator \cite{Alexandru:2011sc}, along with an efficient BiCGstab multi-mass inverter \cite{Alexandru:2011ee} was used to compute all correlation functions in this work.

For the kaon polarizability we needed to include a valence mass for the strange quark. To this end we determined an appropriate value of $\kappa_{\text{s}}$ by measuring the mass of the $\Omega$ baryon and the $\phi$ meson. We computed the mass of the $\Omega$ baryon and $\phi$ meson for a string of $\kappa$ values. We then performed interpolation among these masses to match the physical value of each hadron and took the average of the two values. We find $\kappa_{\text{s}} = 0.1266(1)$  for the $\Ea$ ensemble  and $\kappa_{\text{s}} = 0.1255(1)$ for the $\Ec$ ensemble. 

\subsection{Extracted Parameters}
\label{sec:params}

To reduce our statistical uncertainties we computed quark propagators at multiple point sources for each configuration. The number of sources used in each ensemble is listed in Table~\ref{tab:ensembles}. The sources were chosen by selecting points which are related by translational symmetry. This is achieved by varying the source position only along the $y$ and $z$ axes---these directions remain translationally invariant since we use PBC in these directions.  

To observe the benefits of multiple sources we study the behavior of the statistical error of the energy shift as a function of the number of sources, $N_s$. Fig.~\ref{plot:sources} shows such scaling plots.

In determining the number of sources to compute one should look at the scaling of the statistical error for the different hadrons of interest.  For example, if one were to look only at the pion then it would seem that little or nothing is gained from using more than $N_s \simeq 15$ sources. However, for the neutron we observe a decrease in the uncertainties relative to the pion for up to  25 sources. A similar plateau region for the neutron is expected to occur as more sources are added since the source locations become more dense and hence more correlated.

To quote a value of $\delta E$ we need to determine an appropriate time window to fit the correlators. 
Our fit region is chosen by varying the minimum time distance and computing $ \delta E$, and $\chi^2/\text{dof}$ for each ensemble and particle of interest.  We choose our fit window based on the stability of the parameters as a function of $t_{\text{min}}$, and on a reasonable value of $\chi^2/\text{dof}$.  Fig.~\ref{plot:fitwindow} shows an example of this for the extracted energy shift  of the neutron on the $\Ec$ ensemble at the {\em{unitary}} point, {\em{i.e.}}, the point where the sea quark mass is equal to the valence mass.  The value of $t_{\text{max}}$ was held fixed at  $t = 21$. A similar analysis was performed at neighboring values of $t_{\text{max}}$ and we find the same behavior in each case. For both ensembles and each hadron we determined the fit range using the procedure just described for the unitary point. We then kept that fit window fixed for all other partially-quenched values.
Table~\ref{tab:fitwindow} lists the time ranges used to extract the energy shifts for each hadron on the two ensembles.

\begin{figure*}[!t]
\includegraphics[width= 3.5in]{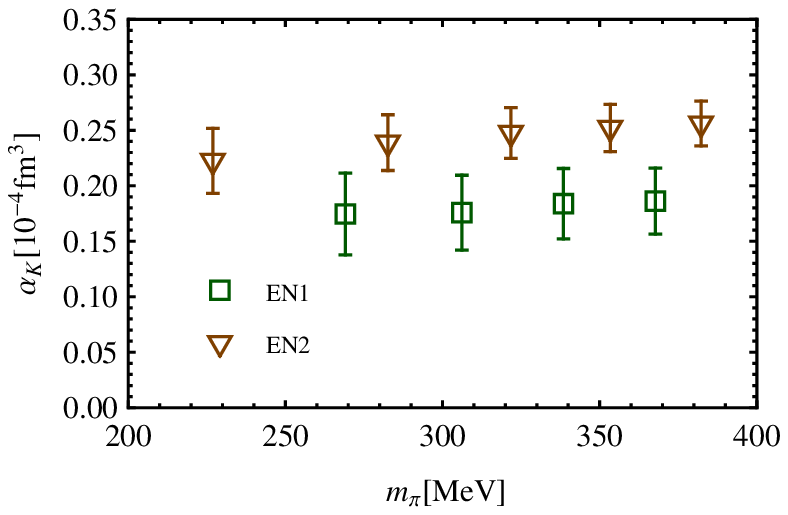}
\includegraphics[width= 3.5in]{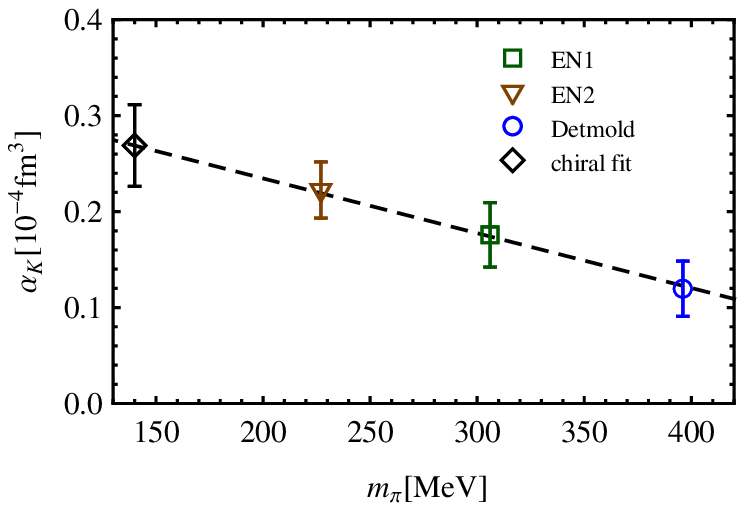}
\caption{Left panel: Plot of the extracted kaon polarizability as a function of the pion mass. Right panel: chiral extrapolation which include only the unitary points for the $\Ea$ and $\Ec$ lattices along with the 400~MeV dynamical point computed in \cite{Detmold:2009dx}.}
\label{plot:kaonpolardynam}
\end{figure*}

\begin{table}[htpb]
\begin{center}
\begin{tabular}{|c|c|c|c|}
 \hline
Ensemble & Pion&Kaon&Neutron\\
\hline
\hline                                                                      
$\Ea$ & [14,30]& [14,30] & [8,21]\\
$\Ec$ & [15,36]& [15,37] & [9,23]\\   
\hline
\end{tabular}
\end{center}
\caption{The list of fit ranges used in extracting the energy shifts for the pion, kaon, and neutron. The fit ranges were determined by
examining the unitary point for each ensemble.}
\label{tab:fitwindow}
\end{table}

\begin{figure}[b]
\includegraphics[width= 3.5in]{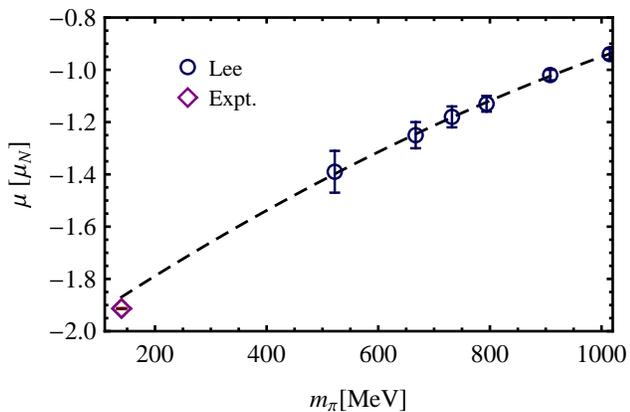}
\caption{Plot of the neutron magnetic moments determined from \cite{Lee:2005ds} as a function of $m_{\pi}$. The black curve is our quadratic fit to the data points.}
\label{plot:magmom}
\end{figure}

\begin{figure*}[t]
\includegraphics[width= 3.5in]{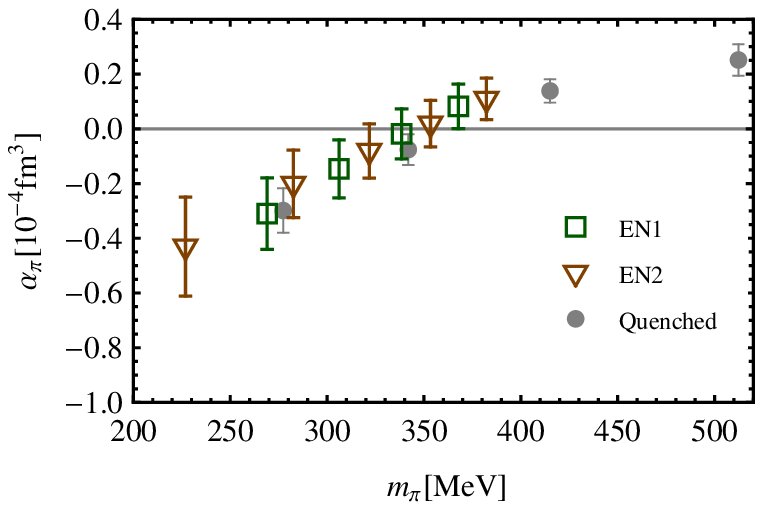}
\includegraphics[width= 3.5in]{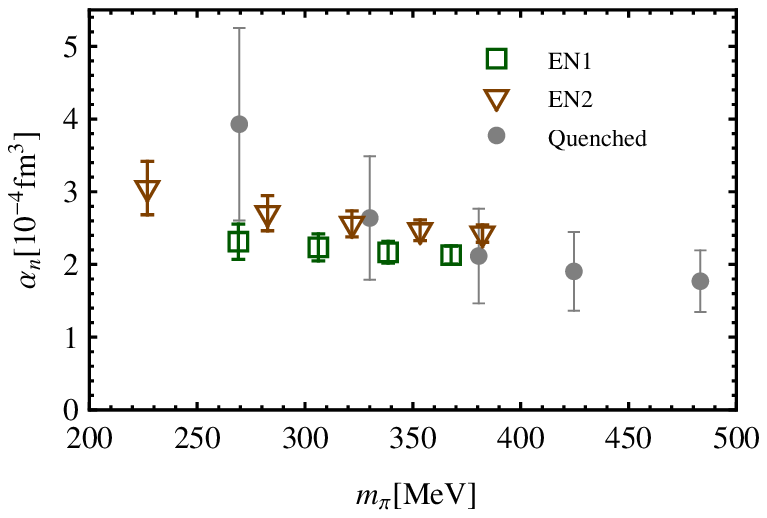}
\caption{Left panel: Plot of the pion polarizability.  Right panel: Our results for the neutron polarizability. The gray/circle points are quenched results found in \cite{Alexandru:2010dx}. }
\label{plot:neutronres}
\end{figure*}

Our computed values for the neutral pion and kaon polarizabilites are presented in Table~\ref{tab:mespolarresults1}. Tables~\ref{tab:mespolarresults2} and \ref{tab:mespolarresults3} list the extracted energy shift and masses respectively.  
The static polarizabilities were computed from the mass shift via the equation:
\begin{equation}
\alpha = \frac{2 a^3 e^2}{9 \eta^2} (a\delta m).
\label{eqnalpha}
\end{equation}
Eq.(\ref{eqnalpha}) is readily obtained by recalling that the mass shift is connected to the polarizability via $\delta m =  \alpha \Ef^2/2$ and $\eta = e a^2 \Ef/3$. The left panel of Fig.~\ref{plot:kaonpolardynam} shows our results for the kaon and the left panel of Fig.~\ref{plot:neutronres} shows our results for the pion, both as a function of $m_{\pi}$. For the pion polarizability we also overlay quenched results from the study done in \cite{Alexandru:2010dx}, we will comment on this in the discussion section.

To obtain the neutron Compton polarizability we use the energy expansion given in  Eq.(\ref{eqn::Eneutron})  which depends on its anomalous magnetic moment $\mu$. 
In our analysis we did not measure $\mu$ directly. Instead we perform an extrapolation to the values of $\mu$ as a function of $m_{\pi}$ determined from an independent study \cite{Lee:2005ds}.  We use a quadratic fit, as was done in \cite{Lee:2005ds}, to find $\mu$ as a function of $m_{\pi}$.
In units of the nuclear magneton, $\mu_{N}$, we find 
\begin{eqnarray}
\label{eqn::mufit}
\mu(m_{\pi}) &=& a_0 + a_1 m_{\pi} + a_2 m_{\pi}^2, \\
\text{where}~&a_0=& -2.067, ~a_1 = 1.459\times 10^{-3}, \nonumber \\ 
\text{and}~&a_2 = &-3.427\times 10^{-7}. \nonumber
\end{eqnarray}
Fig.~\ref{plot:magmom} shows the fit to the data points from  \cite{Lee:2005ds}. 
This fitting form is motivated by $\chi$PT~\cite{Hall:2012pk}. By comparing our fit with
lattice results at lower pion mass~\cite{Primer:2013pva}, we estimate that our systematic
errors are less than 5\%. Given that the magnetic moment contributes at most 20\% to the mass shift the overall systematic associated with this procedure is on the order of 1\%. This is significantly smaller than our stochastic errors on the polarizability.

Using the above functional form for $\mu$ we can now compute the Compton neutron polarizability by
\begin{equation}
\alpha_c = \alpha + \frac{\mu^2(m_{\pi})}{m_n},
\end{equation}
where $\alpha$ is given by Eq.(\ref{eqnalpha}) and $m_n$ is the mass of the neutron measured from the lattice in physical units.

Our extracted parameters for the neutron are also tabulated in Tables~\ref{tab:mespolarresults1}, \ref{tab:mespolarresults2}, and \ref{tab:mespolarresults3}.  In the right panel of Fig.~\ref{plot:neutronres} we plot the results of our two ensembles as a function of $m_{\pi}$. 

\section{Discussion}
\label{sec:discussion}

In this section we will discuss some features in our results for the kaon, pion, and neutron individually. 

\subsection{Neutral Kaon}

Beginning first with the kaon,  we see from the left panel of Fig.~\ref{plot:kaonpolardynam}  that the polarizability  depends on the sea quark mass even when the valence quark mass is kept fixed. We note that the change in the polarizability is small in absolute terms. The difference between the two sea quark masses is only 0.1 in units of $10^{-4}\fm^3$ (the 
``natural" units for hadron polarizability) when the light quark mass is almost halved. The kaon polarizability also changes very little when the valence quark mass is varied. 

This slow change as a function of $m_q$ allows us to do a trustworthy extrapolation to the physical point. To perform this extrapolation, we use the two values at the unitary points and the result determined in \cite{Detmold:2009dx} for the kaon polarizability at $m_{\pi} = 400\MeV$.  The fit assumes a linear dependence on $m_{\pi}$.  The results of our extrapolation are shown in the right panel Fig.~\ref{plot:kaonpolardynam}. We find $\alpha_{\tiny\tiny{K}}= 0.269(43)\times10^{-4} \text{fm}^3$. The kaon polarizability has not yet been measured experimentally. $\chi$PT predicts the polarizability to be zero at $\mathcal{O}(p^4)$~\cite{Guerrero:1997rd}.  Our result is consistent with $\chi$PT since the kaon polarizability is relatively small in units of $10^{-4} \text{fm}^3$.  

There are systematic errors associated with our study. One of them is the tuning of $\kappa_s$. Recall from section~\ref{sec:ensemble} that we tuned the value of $\kappa_s$ so that the masses of the $\Omega$ baryon and the $\phi$ meson matched their physical values. This procedure, however,  produces different kaon masses for the $\Ea$ and $\Ec$ ensembles for comparable light quark masses. This is due to a difference in the strange quark mass on the two ensembles. We do not expect 
this to affect the polarizability significantly. This is supported by the plot in the left panel of Fig.~\ref{plot:kaonpolardynam} where we can see that the polarizability is insensitive to
the value of the valence light quark mass. We expect the same level of insensitivity with respect to the valence strange quark mass.

\subsection{Neutral Pion}

Next we turn to the pion polarizability. In the left panel of Fig.~\ref{plot:neutronres} we overlay the results of our study along with quenched results found in \cite{Alexandru:2010dx}. This comparison among all three data sets tells us that the pion polarizability, at the level of our error bars, is relatively insensitive to the sea quark mass. The quenched ensemble is interpreted as a system where the sea quarks are infinitely heavy. 

We note that the negative trend that has been seen in previous studies~\cite{Detmold:2009dx,Alexandru:2010dx} is still present. We would like to re-iterate what was mentioned in \cite{Detmold:2009dx}: The expectation of $\chi$PT at order $\mathcal{O}(p^4)$ and $\mathcal{O}(p^6)$ is that the $\pi_0$ polarizability is about $\alpha_{\pi_0} \sim -0.5\times10^{-4}\text{fm}^3$ \cite{Portoles:1994rc}. This value is consistent with what we have computed. However, the $\chi$PT results come only from {\em{disconnected}} contributions to the correlation function, which was neglected in our calculation. Without the disconnected contribution it is expected that the polarizability is substantially smaller and positive (see~\cite{Detmold:2009dx}). It was suggested that perhaps this was due to finite-volume effects. However, the study done in \cite{Alexandru:2010dx} and preliminary studies in \cite{Lujan:2013qua} show that this is not the case. This puzzling result could also come from the fact that we have left out the effects of coupling the charge of the sea quarks to the electric field. Different methods to include the effects due to charging of the sea quarks are being explored~\cite{Freeman:2012cy} .

\subsection{The Neutron}

\begin{figure*}[t]
\includegraphics[width= 3.5in]{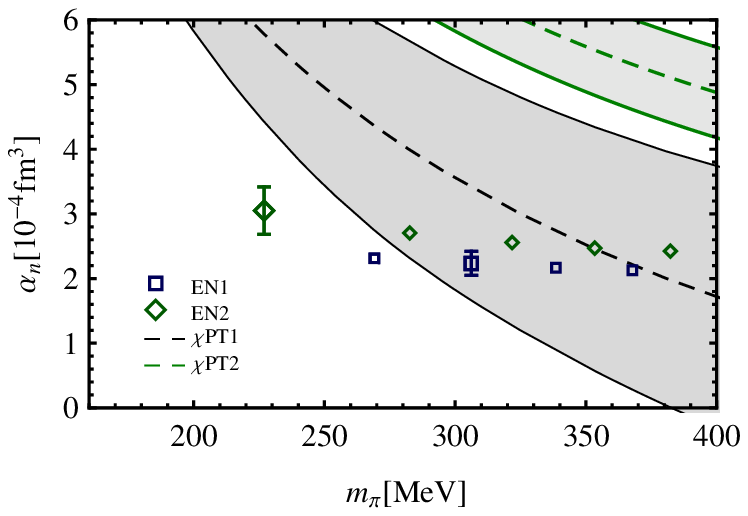}
\includegraphics[width= 3.5in]{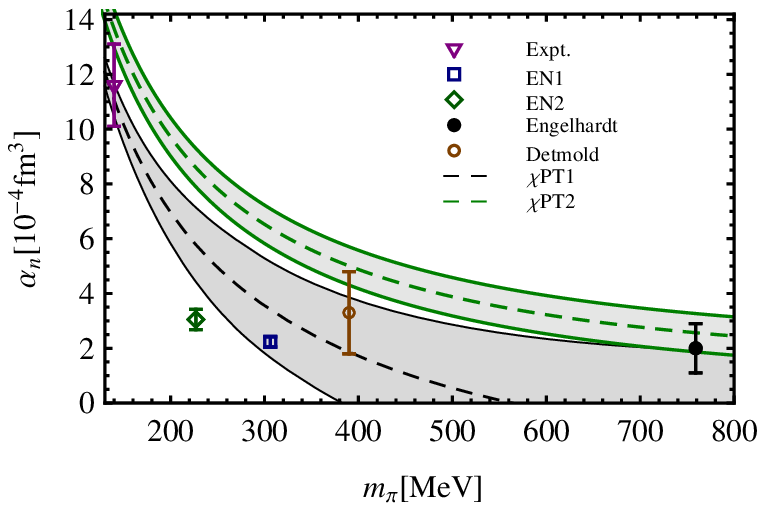}
\caption{Plots of the neutron polarizability as a function of $m_{\pi}$. The left panel shows the results obtained in this work. Only the dynamical points of each ensemble are displayed with error bars. The dashed lines are two different curves predicted by $\chi$PT$1$  \cite{Griesshammer:2012we,McGovern:2012ew} and $\chi$PT$2$ \cite{Lensky:2009uv}.  The right panel plots our data along with the experimental value and two other lattice calculations~\cite{Engelhardt:2007ub} and~\cite{Detmold:2010ts}.}
\label{plot:neutronpolar}
\end{figure*}

In the right panel of Fig.~\ref{plot:neutronres} we plot our results for neutron polarizability. We see that the
dependence on the sea quark mass is more pronounced than in the meson case. Our results are compatible with the
quenched ones, at least qualitatively. The large error bars for the quenched results make a more quantitative
evaluation impossible. Turning to our results, we see that the polarizability rises when the quark mass is
decreased, as anticipated. The change in the valence mass produces only a slight increase, whereas the sea quark
mass change plays a more important role.

We compare now our findings for the neutron polarizability to two different $\chi$PT predictions: $\chi$PT$1$~\cite{Griesshammer:2012we,McGovern:2012ew} and $\chi$PT$2$~\cite{Lensky:2009uv}. 
We use this comparison to gauge the systematic errors of our calculation, in particular
finite volume effects and neglecting the electric charge of the sea quarks.
These two $\chi$PT curves use different approximations in their calculations to derive the chiral form. In the case of $\chi$PT$1$ the calculation is expanded to N$^2$LO using a non-relativistic form for the propagators. There are two extra free parameters  which are determined by fitting to Compton scattering data. The second result, $\chi$PT$2$, includes terms up to NLO and uses relativistic propagators. They compute $\alpha$ as a function of $m_{\pi}$ with no free parameters. The error bars in $\chi$PT$1$ come from a careful analysis \cite{Griesshammer:inprep} whereas the error bar for the second curve is fixed to a value estimated at the physical point.

The left panel of Fig.~\ref{plot:neutronpolar} shows the two $\chi$PT curves along with our findings. Our results for both ensembles seem to be in agreement more with the $\chi$PT$1$ curve for our 306 MeV pion.  However, our lattice calculation is in disagreement with both curves at the 227 MeV pion.  We believe that this is due to finite-volume effects and the fact that the sea quarks are electrically neutral. 

To gauge the effect of charging the sea quarks we use $\chi$PT~\cite{PhysRevD.73.114505}: 
for the pion mass between $140\MeV$ and $300\MeV$ when the sea quarks are charged
the polarizability increases by
$1.5$--$2\times 10^{-4}\fm^3$. This would explain part of the discrepancy seen between our
data at $m_\pi=227\MeV$ and the $\chi$PT curves shown in Fig.~\ref{plot:neutronpolar}. 
However, significant
differences still remain and we believe that this is due to finite volume effects.

Finite volume corrections can also be estimated using $\chi$PT.  For periodic boundary conditions these effects were calculated for electric polarizabilities~\cite{PhysRevD.73.114505} and magnetic polarizabilities \cite{Hall:2013dva}. For $m_{\pi} = 250$ MeV and $L=3 \fm$ it was found that the correction to $\alpha$ is about 7\%~\cite{PhysRevD.73.114505}.  However, we used DBC in this work and we expect that these corrections will be more important than for PBC. This is supported by the discrepancy we have between our results and $\chi$PT predictions, as discussed above, and sigma model studies for chiral condensate in the presence of hard walls~\cite{Tiburzi:2013vza}. Further studies are required to determine the magnitude of these corrections.

On the right panel of Fig.~\ref{plot:neutronpolar} we add the experimental point along with two other lattice calculations~\cite{Detmold:2010ts,Engelhardt:2007ub} for the neutron polarizability. Our results have the smallest pion masses used in polarizability studies and the smallest statistical errors. 

\section{Conclusion}
\label{sec:conclusion}

We performed a valence calculation of the electric polarizabilites of the neutral pion, neutral kaon, and neutron using a two-flavor nHYP clover action at two dynamical pion masses: 306 MeV and 227 MeV. These are to date the lowest pion masses used for polarizability studies.  A chiral extrapolation for the kaon was performed using three dynamical points including the 400 MeV point from \cite{Detmold:2009dx}. We find the neutral kaon polarizability to be $\alpha_{\tiny\tiny{K}} = 0.269(43)10^{-4}\text{fm}^3$. The chiral behavior of the kaon is fairly mild, suggesting that the systematic errors for our extrapolated value are similarly mild. For the pion, the negative trend remains to be understood. We speculate that this may be due to the fact that we have neglected the charge of the sea quarks and we are working on including these effects \cite{Freeman:2012cy}. Our neutron polarizability results are promising. The stochastic errors are significantly smaller than other lattice studies and we also have the lightest dynamical quark masses. We note that our errors are significantly smaller than the ones from $\chi$PT studies. The hope is that when the finite-volume systematics are removed and the sea quarks are charged we will be able to constrain the parameters in the $\chi$PT models. This in turn could be used to tighten the error bars on the $\chi$PT predictions at the physical point and make the comparison with the experimentally measured values more informative.

\begin{acknowledgements}

We would like to thank Craig Pelissier for generating the ensembles used in this study. Also many thanks to Harald Grie\ss hammer and Vladimir Pascalutsa  for providing us with the $\chi$PT curves for the neutron polarizability. The computations were carried out on a variety of GPU-based platforms, including the IMPACT clusters and  Colonial One at GWU,  the clusters at JLab and FermiLab, and the clusters at University of Kentucky.  This work is supported in part by the NSF CAREER grant PHY-1151648, the U.S. Department of Energy grant DE-FG02-95ER40907, and the ARCS foundation.

\end{acknowledgements}

\bibliographystyle{jhep-3}
\bibliography{my-references}

\begin{table*}[b!]
\begin{tabular}{|c|c|c|c|c|c|c|c|c|c|c|}
\hline
\multicolumn{1}{|c}{}& \multicolumn{10}{|c|}{ $\alpha$ [$10^{-4} \text{fm}^3$]~~~~~~~~~~~~~~~  }\\
\cline{2-11}
\multicolumn{1}{|c}{Hadron}& \multicolumn{4}{|c|}{ $\Ea$ }& \multicolumn{1}{}{c}&\multicolumn{5}{|c|}{ $\Ec$ }\\
\cline{2-5}
\cline{7-11}
                    &$m_1$&$m_2$& $m_3$&$m_4$ &  &$m_1$&$m_2$ & $m_3$ & $m_4$& $m_5$ \\
\cline{1-5} \cline{7-11}
    \multirow{1}*{$\pi$} &$-0.31(13)$ & $-0.15(11)$& $-0.019(91)$ &$0.082(81)$&  &$-0.43(18)$& $-0.20(12)$ & $-0.082(0.10)$ &$0.019(85)$ & $0.109(76)$\\
                                      
                                        \cline{1-5} \cline{7-11}
   \multirow{1}*{$K$}    &$0.175(37)$  &$0.176(34)$ & $0.184(32)$ & $0.186(30)$&  &$0.222(29)$  &$0.239(25)$  & $0.248(23)$ & $0.252(21)$ & $0.256(20)$\\
                                      
                                       \cline{1-5} \cline{7-11}
   \multirow{1}*{$n$}    &$2.31(24)$ &$2.24(18)$&$2.17(15)$&$2.13 (13)$& &$3.06(0.37)$ &$2.71(24)$&$2.56(18)$&$2.48(14)$&$2.43(12)$\\     
                                                                                                       
\hline
\end{tabular}
\caption{Extracted polarizabilities for both ensembles and for all kappa values.  The labels $m_1, m_2$, etc. correspond to the pion masses that are tabulated in Table~\ref{tab:partialquench}.}
\label{tab:mespolarresults1}
  \end{table*}

\begin{table*}[t]
\begin{tabular}{|c|c|c|c|c|c|c|c|c|c|c|}
\hline
\multicolumn{1}{|c}{}& \multicolumn{10}{|c|}{ $a\delta E \times 10^{8}$~~~~~~~~~~~~~~~ }\\
\cline{2-11}
\multicolumn{1}{|c}{Hadron}& \multicolumn{4}{|c|}{ $\Ea$ }& \multicolumn{1}{}{c}&\multicolumn{5}{|c|}{ $\Ec$ }\\
\cline{2-5}
\cline{7-11}
                    &$m_1$&$m_2$& $m_3$&$m_4$ &  &$m_1$&$m_2$ & $m_3$ & $m_4$& $m_5$ \\
\cline{1-5}
\cline{7-11}
    \multirow{1}*{$\pi$} &$-7.37(3.10)$ &$-3.69(2.68)$ &$-0.49(2.39)$ &$2.20(2.18)$& &$-10.01(4.24)$&$-5.26(3.23)$  &$-2.24(2.73)$ &$0.54(2.44)$&$3.21(2.22)$\\
                                        
                                       \cline{1-5} \cline{7-11}
   \multirow{1}*{$K$}    &$5.08(1.07)$&$5.14(99)$ &$5.39(93)$ &$5.48(88)$& &$7.14(94)$ &$7.69(81)$ &$7.99(74)$ &$8.15(69)$ &$8.29(65)$\\
                                        
                                          \cline{1-5} \cline{7-11}
   \multirow{1}*{$n$}   & $55.1(7.2)$ &$53.4(5.4)$ &$52.2(4.4)$ &$51.8(3.67)$& &$86.2(12.5)$ &$74.5(7.9)$&$70.7(5.8)$ &$68.6(4.7)$ &$67.9(3.8)$\\
                                                                                                       
\hline
\end{tabular}
\caption{Extracted energy shifts for both ensembles and for all kappa values.  The labels $m_1, m_2$, etc. correspond to the pion masses that are tabulated in Table~\ref{tab:partialquench}.}
\label{tab:mespolarresults2}
  \end{table*}

\begin{table*}[htpb]
\begin{tabular}{|c|c|c|c|c|c|c|c|c|c|c|}
\hline
\multicolumn{1}{|c}{}& \multicolumn{10}{|c|}{ $am$~~~~~~~~~~~~~~~ }\\
\cline{2-11}
\multicolumn{1}{|c}{Hadron}& \multicolumn{4}{|c|}{ $\Ea$ }& \multicolumn{1}{}{c}&\multicolumn{5}{|c|}{ $\Ec$ }\\
\cline{2-5}
\cline{7-11}
                    &$m_1$&$m_2$& $m_3$&$m_4$ &  &$m_1$&$m_2$ & $m_3$ & $m_4$& $m_5$ \\
\cline{1-5} \cline{7-11}  \multirow{1}*{$K$} &$0.3155(7)$&$0.3220(7)$&$0.3282(6)$ &$0.3344(6)$& & $0.3698(6)$&$0.3767(6)$&$0.3825(5)$ &$0.3878(5)$&$0.3930(5)$ \\
                                     
                                        \hline
   \multirow{1}*{$n$}   &$0.632(8)$&$0.644(6)$&$0.659(5)$&$0.674(4)$& &$0.618(13)$&$0.628(8)$&$0.643(6)$&$0.658(5)$ &$0.672(4)$ \\    
                                                                                                        
\hline
\end{tabular}
\caption{Masses of the kaon and neutron for both ensembles and for all kappa values. The labels $m_1, m_2$, etc. correspond to the pion masses that are tabulated in Table~\ref{tab:partialquench}.}
\label{tab:mespolarresults3}
  \end{table*}

\end{document}